\documentclass[a4paper, 12pt]{article} 
\usepackage[right=1.1in,left=1.1in,top=1.1in,bottom=1.1in]{geometry}
\usepackage{graphicx} 
\usepackage{braket}
\usepackage{mathrsfs}
\usepackage[T1]{fontenc} 
\usepackage{amsmath}
\usepackage{pxfonts}
\usepackage[T1]{fontenc}
\usepackage{amssymb}
\usepackage{natbib}
\usepackage{epstopdf}
\usepackage{setspace}
\usepackage{enumitem}
\usepackage{sectsty}
\usepackage{wrapfig} 
\sectionfont{\large}
\bibpunct{(}{)}{,}{a}{}{,}
\usepackage{tikz}   
\usepackage{tikz-3dplot} 

\newcommand{\qed}{\nobreak \ifvmode \relax \else
      \ifdim\lastskip<1.5em \hskip-\lastskip
      \hskip1.5em plus0em minus0.5em \fi \nobreak
      \vrule height0.75em width0.5em depth0.25em\fi}

\usepackage{bbm}
\usepackage{MnSymbol}
\usepackage[all]{xy}

\usepackage{xcolor,colortbl,array,amssymb}

\makeatletter

\title{\large{Review of \emph{The Quantum Revolution in Philosophy}, by Richard Healey, Oxford: Oxford University Press, 2017. }} 

\author{Eddy Keming Chen\thanks{Department of Philosophy,  University of California, San Diego, 9500 Gilman Dr, La Jolla, CA 92093. Website: www.eddykemingchen.net. Email: eddykemingchen@ucsd.edu  }}

\date{\emph{The Philosophical Review} (2020) 129 (2): 302-308 } 

\begin{document}
\bibliographystyle{plain}

\maketitle 







\nocite{einstein1935can, healey2017quantum}

In this thought-provoking book, Richard Healey proposes a new interpretation of quantum theory inspired by pragmatist philosophy. Healey puts forward the interpretation as an alternative to realist quantum theories on the one hand such as Bohmian mechanics, spontaneous collapse theories, and many-worlds interpretations, which are different proposals for describing what the quantum world is like and what the basic laws of physics are, and non-realist interpretations on the other hand such as quantum Bayesianism, which proposes to understand quantum theory as describing agents'  \emph{subjective} epistemic states. The central idea of Healey's proposal is to understand  quantum theory as providing \emph{not} a description of the physical world but a set of authoritative and \emph{objectively} correct prescriptions about how agents should act. The book provides a detailed development and defense of that idea, and it contains interesting discussions about a wide range of philosophical issues such as representation, probability, explanation, causation, objectivity, meaning, and fundamentality. 

Healey's project is at the intersection of physics and philosophy. The book is divided into two parts. Part I of the book  discusses  the foundational questions in quantum theory from the perspective of the prescriptive interpretation. In Part II, Healey discusses the philosophical implications of the view. Both parts are written in a way that is largely accessible to non-specialists. 
 In this brief book review, I will focus on two  questions:

\begin{itemize}
	\item[(1)] How does Healey's idea work?
	\item[(2)] What reasons are there to believe in it? 
\end{itemize}


To understand Healey's ideas, we should provide some background for non-specialists. Quantum theory has been subject to intense scrutiny and philosophical investigations in the last century. There has been enormous progress on several fronts, including the empirical tests about Bell inequalities and the development  of several realist interpretations. It has been proposed that what is revolutionary about quantum theory and what sets it apart from classical physics are quantum entanglement and quantum non-locality. For many realists, it seems that quantum theory  requires us to postulate a different kind of ontology and a different kind of laws of nature. 

Quantum entanglement describes the holistic phenomena that there are physical properties that can be attributed to a composite system that are not the product of properties of its  parts. For example, two electrons $A$ and $B$ are entangled when the pair has a quantum state that is not the product of the state of $A$ and the state of $B$. Quantum entanglement is reflected in the way quantum states, which are typically represented by ``wave functions,'' get assigned to different systems.  

Quantum non-locality (also called ``spooky action at a distance'') describes the phenomena that events that are arbitrarily far apart can influence each other and the strength of the influence does not depend on their distance. Quantum non-locality is a strange feature, and its falsity was assumed in the  Einstein-Podolsky-Rosen (1935) argument for the incompleteness of quantum theory.  Remarkably, \cite{bell1964einstein} proved  that if the quantum predictions are accurate, then nature is non-local. Subsequent experimental tests since the 1980s have verified the quantum predictions to a high level of precision, reinforcing the conclusion that nature is non-local.\footnote{ See \cite{maudlin2011quantum} for a detailed discussion of these issues.}

Both Bohmian mechanics and spontaneous collapse theories are non-local. And all three realist theories affirm the reality of quantum entanglement. The reality of quantum entanglement follows from the reality of the quantum state. After all, the quantum state is so successful in application. If one wants to be a realist, it seems that the most natural way is to accept the reality of the quantum state and its consequences.  But what is represented by the wave function? What is its nature? How does it relate to things in the physical world? A hot topic in recent work in philosophy of physics concerns exactly those questions.\footnote{See \cite{chen2019realism} for an overview.}

I think the key to understand Healey's idea is to think of it as a proposal for the proper foundation of quantum theory---a proposal to replace descriptive facts with normative facts as the foundation, and to replace ``what there is'' with ``what we ought to expect.''  Healey's approach would  sound alien to many people, since it is not realist in any traditional sense. It is common to think that fundamental physics, and in particular quantum theory, is in the business of saying what the world is like (even though it may be difficult to discern what it says). It is true that we can learn useful advice from a correct physical theory, such as:

\begin{itemize}
	\item One should avoid falling into a black hole.
	\item One should place  detectors far from sources of interference.
	\item One should get away from a nuclear explosion. 
	\item One should be almost certain to see this particle decay in 1 month. 
\end{itemize}
 A correct physical theory may explain the correctness of useful practical advices such as above. 
Such advices are \emph{objectively} correct in so far as they are derived from the conjunction of some accurate descriptions of physical reality and some correct theories of human rationality.\footnote{ For example, one should avoid dangerous activities, minimize errors,  and maximize epistemic accuracy.} For Healey, that might be a good interpretation of  classical physics, but that is not the right interpretation of quantum theory. Quantum theory, for Healey, directly offers authoritative advice about what we should believe and expect, and such advice is not derived in any way from some underlying descriptive facts. Still, such advice can be objectively true (unlike the quantum Bayesian's emphasis on subjective epistemic states).  The advice has the form ``one should assign quantum state $\psi$ to the physical system $s$,'' and ``one should have $0.5$ confidence in  $M$,'' where $M$ is some kind of magnitude claim about the physical system $s$. 
On Healey's interpretation, quantum theory is about such advice, not some descriptions about what nature is like. Although the magnitude claims are descriptions of nature, they are not contained in quantum theory. Instead, they are assumed to be the targets of explanation, and they are somewhat prior to the application of quantum theory. However, they will be suitably constrained by quantum theory, and only a subclass of them will be \emph{empirically significant.}  

There are many parts to Healey's proposal. To simplify our discussions below,  here is  a short summary of the key points (mostly from Chapter 5):

\begin{itemize}
	
\item Advice on state assignments: agents should follow the standard practice in quantum physics and assign  an appropriate quantum state (e.g. a wave function $\psi$) to a system relative to their physical situation. 

\item Magnitude claims:  a canonical magnitude claim has the form ``the value of dynamical variable $M$ on physical system $s$ lies in set $\Delta$.'' One example would be ``the position of the particle is on the left side of the box.'' Magnitude claims are assumed in the theory and they are the targets of explanation. (See \S5.2 and \S12.3 for elaborations.)

\item Empirical significance: a magnitude claim is empirically significant  if the relevant quantum state the agent should assign to the system has properly experienced \emph{decoherence} due to interactions with the environment, where decoherence is a mathematical property of the quantum state. (See \S5.2 and Appendix C for more on decoherence).

 	\item Advice on credences: agents should adopt credences about \emph{empirically significant} magnitude claims according to the Born rule (setting one's credence about an empirically significant magnitude claim to be  the squared modulo of the appropriate wave function). 

  	\item Descriptions of nature: none of the above, except for the magnitude claims, describes anything in nature. However, magnitude claims are not contained or entailed by quantum theory. They are given prior to the theory. 

\end{itemize}
Let us walk through a simple example. Suppose our physicist Alice sets up an experiment for the position measurement of a particle $s$. Alice wants to explain some magnitude claim about the particle's position (say that it is on the left side of the box). Quantum theory tells her that she ought to assign a particular wave function $\psi$ to the particle. (The magnitude claim about the particle's position will only become empirically significant when the particle has appropriately interacted with the environment such that  different parts of the wave function are almost separated into macroscopically distinct regions as described by decoherence.) Her expectation about the (empirically significant) magnitude claim should match the probability given by the Born rule, which is the squared modulo of the wave function at this moment. 

Many questions  arise at this point. For example: what makes a particular quantum state assignment \emph{appropriate}? It is standard to assume that a  state assignment is appropriate or correct in virtue of  representational accuracy, that is, the system indeed has the particular quantum state. But Healey disagrees; it is more or less a basic fact and if we want to explain it we can appeal to the practice within the scientific community, not some metaphysical facts about what nature is like. It seems to me that on Healey's project the correctness of state assignments is a fundamental normative fact, and any further inquiry about it may be a meta-normative project. This sheds light on how Healey tries to solve (or dissolve) traditional problems in quantum foundation. In the following, we discuss how Healey applies his interpretation to the quantum measurement problem, quantum entanglement, and Bell non-locality.

The quantum measurement problem is often presented as an inconsistency among three propositions: (1) the quantum state is a complete description of nature; (2) the quantum state always changes according to some unitary equation (e.g. the Schr\"odinger equation); (3) experiments have determinate (unique) outcomes.  Applying this to the above experimental set up  involving Alice, we can derive a contradiction. If the quantum state is complete and it changes unitarily, then quantum theory implies there is no determinate outcome of the position measurement of the particle, since it will always be in a \emph{superposition} of $\psi_{\text{left}}$ and $\psi_{\text{right}}$, which is inconsistent with (3)---the particle is measured to be either on the left side or the right side of the box. For Healey, (1) and (2) are simply non-starters, and (3) is obviously true, which means that there is no problem to begin with. (See \S6 and especially p.99.) Quantum theory is prescriptive and it is mistaken to think that the quantum state describes the world, much less describing it completely. Moreover, in typical measurement contexts, such as in Alice's case, the appropriate quantum state has decohered after measurement, giving Alice permission to apply the Born rule to make probabilistic predictions. There is a determinate outcome of measurement, but it is not entailed or contained in quantum theory. Instead, determinate outcomes are something confirmed by our direct experiences. Furthermore, if the outcome were not determinate, quantum theory would lose  its empirical justification. So in order to apply quantum theory, we have to \emph{assume} that experimental outcomes are unique and determinate (in typical cases). This stands in sharp contrast to other single-world theories such as Bohmian mechanics and spontaneous collapse theories that aim to justify (3) from a physical point of view. They seek to derive (3) as a theorem, by applying the precise mathematical equations to the microscopic ontology. 

For Healey, quantum entanglement is no longer metaphysically revolutionary. State assignments may be entangled and non-separable  in the sense that the complete state of the system does not supervene on the states of its parts. However, it seems that Healey would say that the supervenience has no metaphysical significance but only normative significance. The advice  about which state to assign to  the whole system does not supervene, in general, on any advice on what to assign to the parts.  

Healey thinks that Bell non-locality becomes more digestible on the prescriptive view (\S4). Healey does not dispute Bell's proof as a demonstration of the existence of non-local correlations, i.e. probabilities of events happening in region $A$ may depend on events happening in region $B$ which can be arbitrarily far away from $A$. But Healey does not think non-local correlations imply non-local influences or causation. What Bell's theorem (and the subsequent empirical tests) really shows is that the quantum probability of an event localized in some region must be ``relativized to arbitrarily distant places and times.'' But why should we think such a thing is true? It is standard to infer there is some goings-on in the world that accounts for the non-local correlations---some kind of underlying physical mechanism of non-local influence. For example, the mechanism  is accounted for, in explicit physical details, by the guidance equation in Bohmian mechanics and the collapse mechanism in spontaneous collapse theories.  But that is a mistake from Healey's perspective. Quantum theory postulates such non-local correlations as  reliable prescriptions for agents who want to use quantum theory successfully, and it is a mistake to try to describe in virtue of what do such correlations obtain.

What are the reasons for accepting Healey's new interpretation of quantum theory? I think the main reason, for Healey, is that it allows us to avoid the surprising metaphysical consequences such as non-locality and the reality of the quantum state. But at what costs?  If Healey is right, then quantum theory is no longer in the business of describing nature. For most people, I think, that would be too costly. 

What about  alternatives to Healey's theory? How do they compare to it?  Healey gives reasons for thinking that they are inadequate (\S7). He thinks that spontaneous collapse theories are too speculative, and that they do not qualify as ``interpretations.'' He suggests that many-worlds theories still face the conceptual problems of probability and emergence of macroscopic branches. He notes that Bohmian mechanics faces problems in relativistic contexts. Interestingly, Healey's objection to Bohmian mechanics is that it would postulate theoretical structure (e.g. preferred foliation of space-time, particle motion, and the universal wave function)  that is empirically inaccessible.   Healey does not discuss whether these are more costly than abandoning the descriptive role of quantum theory, but I think he would say so.  It is important to note that all of them are not only candidate descriptions of nature but are also in principle applicable to the universe as a whole, while Healey's interpretation would not be, since on his theory, the application of Born rule is only appropriate when there is environmental decoherence (and the universe as a whole by definition has no external environment to interact with) (p. 227).\footnote{For a detailed and updated  introduction to  the  realist theories that are alternatives to Healey's, see \cite{norsen2017}.}

As is the case with Healey's previous books, this one contains many interesting and controversial ideas, which will generate much debate in the years to come. Above all, Healey should be applauded for writing a clear, accessible, and engaging introduction to  the prescriptive interpretation of quantum theory. There are many issues that Healey discusses at length in the book that I do not have space to comment on. Even though we may not agree on the \emph{truth} of the matter, we can certainly agree that Healey's book will be \emph{useful} to anyone interested in the intersection of physics and philosophy.




\bibliography{test}


\end{document}